\documentclass[conference,compsoc,nofonttune]{IEEEtran}
\usepackage[nocompress]{cite}
\usepackage[utf8]{inputenc}
\usepackage{microtype}
\usepackage[ruled,vlined]{algorithm2e}
\usepackage{tikz}
\usepackage{booktabs}
\usepackage{amsmath}
\usepackage{amssymb}
\usepackage{mathtools}
\usepackage{interval}
\usepackage[obeyspaces,spaces]{url}

\usetikzlibrary{positioning}
\tikzset{
	switch/.style={
		rectangle,draw,inner sep=0pt,
		minimum height=3mm,minimum width=6mm
	},
	leaf/.style={
		rectangle,draw,fill=gray,inner sep=0pt,
		minimum height=3mm,minimum width=6mm
	},
	NIC/.style={
		circle,draw,inner sep=0pt,minimum size=3mm
	},
}
\def\assign{\leftarrow}
\def\connabove{\rotatebox[origin=c]{270}{$\Rsh$}}
\def\connbelow{\rotatebox[origin=c]{270}{$\Lsh$}}
\mathchardef\period=\mathcode`.
\DeclareMathSymbol{.}{\mathord}{letters}{"3B} % chktex 18

\begin{document}

\title{High-Quality Fault-Resiliency in Fat-Tree Networks
	\textit{(extended abstract)}}

\author{%
	\IEEEauthorblockN{%
		John Gliksberg%
		\IEEEauthorrefmark{1}\IEEEauthorrefmark{2}\IEEEauthorrefmark{3},
		Antoine Capra\IEEEauthorrefmark{2},
		Alexandre Louvet\IEEEauthorrefmark{2},
		Pedro Javier García\IEEEauthorrefmark{3}, and
		Devan Sohier\IEEEauthorrefmark{1}
	}\vspace{6pt}
	\IEEEauthorblockA{%
		\begin{tabular}{r@{\hskip 6pt}lr@{\hskip 4pt}l}
		\IEEEauthorrefmark{1}
		Li-Parad, UVSQ,& Versailles, France,\quad&
		\{\texttt{john.gliksberg}, \texttt{devan.sohier}\}
		&\texttt{@uvsq.fr} \\
		\IEEEauthorrefmark{2}
		BXI, Atos,& Bruyères-le-Châtel, France,\quad&
		\{\texttt{antoine.capra}, \texttt{alexandre.louvet}\}
		%\texttt{antoine.capra}
		&\texttt{@atos.net} \\
		\IEEEauthorrefmark{3}
		RAAP, UCLM,& Albacete, Spain,\quad&
		\texttt{pedrojavier.garcia}&\texttt{@uclm.es}
		\end{tabular}
	}
}

\maketitle

\begin{abstract}
	Coupling regular topologies with optimized routing algorithms
	is key in pushing the performance
	of interconnection networks of HPC systems.
	In this paper we present Dmodc,
	a fast deterministic routing algorithm
	for Parallel Generalized Fat-Trees (PGFTs)
	which minimizes congestion risk
	even under massive topology degradation caused by equipment failure.
	It applies a modulo-based computation of forwarding tables
	among switches closer to the destination,
	using only knowledge of subtrees for pre-modulo division.
	Dmodc allows complete re-routing of topologies
	with tens of thousands of nodes in less than a second,
	which greatly helps centralized fabric management
	react to faults with high-quality routing tables
	and no impact to running applications
	in current and future very large-scale HPC clusters.
	%(This is already the case in an 8490-node production cluster.)
	We compare Dmodc against routing algorithms
	available in the InfiniBand control software (OpenSM)
	first for routing execution time to show feasibility at scale,
	and then for congestion risk under degradation
	to demonstrate robustness.
	The latter comparison is done using static analysis
	of routing tables under random permutation (RP),
	shift permutation (SP) and all-to-all (A2A) traffic patterns.
	Results for Dmodc show A2A and RP congestion risks
	similar under heavy degradation
	as the most stable algorithms compared,
	and near-optimal SP congestion risk
	up to 1\% of random degradation.
	%Finally, we provide a node-type-specific variant (gDmodc)
	%and conduct similar analysis with node-type-specific traffic
	%to demonstrate improved results
	%provided the traffic hypothesis holds.
\end{abstract}

%\begin{IEEEkeywords}
%    HPC, routing, fat-tree, fault-resiliency%, heterogeneity
%\end{IEEEkeywords}

\section{Introduction}

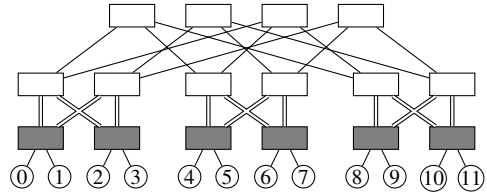
\begin{figure}
\centering
\begin{tikzpicture}
    \node[leaf]   (L1-0)                     {};
    \node[leaf]   (L1-1) [right=4mm of L1-0] {};
    \node[leaf]   (L1-2) [right=6mm of L1-1] {};
    \node[leaf]   (L1-3) [right=4mm of L1-2] {};
    \node[leaf]   (L1-4) [right=6mm of L1-3] {};
    \node[leaf]   (L1-5) [right=4mm of L1-4] {};
    \node[switch] (L2-0) [above=4mm of L1-0] {};
    \node[switch] (L2-1) [above=4mm of L1-1] {};
    \node[switch] (L2-2) [above=4mm of L1-2] {};
    \node[switch] (L2-3) [above=4mm of L1-3] {};
    \node[switch] (L2-4) [above=4mm of L1-4] {};
    \node[switch] (L2-5) [above=4mm of L1-5] {};
    \node[switch] (L3-1) [above=6mm of L2-2] {};
    \node[switch] (L3-2) [above=6mm of L2-3] {};
    \node[switch] (L3-0) [left=4mm  of L3-1] {};
    \node[switch] (L3-3) [right=4mm of L3-2] {};
    \node[NIC] (nic0)  [below=2mm of L1-0,xshift=-2.5mm] {\footnotesize  0} edge (L1-0);
    \node[NIC] (nic1)  [below=2mm of L1-0,xshift=+2.5mm] {\footnotesize  1} edge (L1-0);
    \node[NIC] (nic2)  [below=2mm of L1-1,xshift=-2.5mm] {\footnotesize  2} edge (L1-1);
    \node[NIC] (nic3)  [below=2mm of L1-1,xshift=+2.5mm] {\footnotesize  3} edge (L1-1);
    \node[NIC] (nic4)  [below=2mm of L1-2,xshift=-2.5mm] {\footnotesize  4} edge (L1-2);
    \node[NIC] (nic5)  [below=2mm of L1-2,xshift=+2.5mm] {\footnotesize  5} edge (L1-2);
    \node[NIC] (nic6)  [below=2mm of L1-3,xshift=-2.5mm] {\footnotesize  6} edge (L1-3);
    \node[NIC] (nic7)  [below=2mm of L1-3,xshift=+2.5mm] {\footnotesize  7} edge (L1-3);
    \node[NIC] (nic8)  [below=2mm of L1-4,xshift=-2.5mm] {\footnotesize  8} edge (L1-4);
    \node[NIC] (nic9)  [below=2mm of L1-4,xshift=+2.5mm] {\footnotesize  9} edge (L1-4);
    \node[NIC] (nic10) [below=2mm of L1-5,xshift=-2.5mm] {\footnotesize 10} edge (L1-5);
    \node[NIC] (nic11) [below=2mm of L1-5,xshift=+2.5mm] {\footnotesize 11} edge (L1-5);
    \draw[-,double distance=0.3mm] (L1-0) to (L2-0) to (L1-1) to (L2-1) to (L1-0);
    \draw[-,double distance=0.3mm] (L1-2) to (L2-2) to (L1-3) to (L2-3) to (L1-2);
    \draw[-,double distance=0.3mm] (L1-4) to (L2-4) to (L1-5) to (L2-5) to (L1-4);
    \draw[-] (L3-0) to (L2-0) to (L3-2) to (L2-2) to (L3-0) to (L2-4) to (L3-2);
    \draw[-] (L3-1) to (L2-1) to (L3-3) to (L2-3) to (L3-1) to (L2-5) to (L3-3);
\end{tikzpicture}
	\caption{$PGFT(3;2.2.3;1.2.2;1.2.1)$
		with leaf switches shown in grey.}\label{fig:pgft}
\end{figure}

%\subsection*{State of the art}
%
A majority of current leading network topologies
for High Performance Computing (HPC) clusters
are fat-tree variants.
(The five most powerful clusters of
the November 2018 Top500 list\cite{top500}
%(According to the Top500 list\cite{top500},
%the five most powerful clusters of November 2018
boasted fat-tree topologies.)
It is sufficient for
fat-tree-specific routing algorithms
to be minimal to guarantee deadlock-free routing,
%only need to be minimal to guarantee deadlock-free routing,
%need only be minimal to guarantee deadlock-free routing,
and the regular nature of their target topology class
should simplify load-balancing strategies.
%Various versions have been described in many papers:
%Leiserson's algorithm for the original fat-tree design\cite{leiserson1985fat}
%proves the existence of near-optimal offline schedulings
%of message sets into delivery frames
%(since the scheduler did not directly control channel selection,
%rather tried to minimize packet dropping),
%while algorithms designed for fat-tree variants
%Unlike Leiserson's original routing\cite{leiserson1985fat},
%algorithms designed for fat-tree variants
%based on folded Clos networks
%optimize path selection to minimize congestion.
In general, oblivious routing
(without knowledge of communication patterns)
in fat-trees is deterministic
and optimized for shift patterns~\cite{gomez2007deterministic}%
\cite{zahavi2010optimized}\cite{zahavi2014quasi}.
In particular, PGFTs~\cite{zahavi2010d}
describe all regular fat-trees for which there is at most
one downward switch-path from any switch to any node
(as shown in Figure~\ref{fig:pgft}).
The oblivious algorithm for non-degraded PGFTs (Dmodk)
%their corresponding oblivious non-fault-resilient algorithm Dmodk
uses this property and their connection logic
to provide load balance through an arithmetic rule.
%Dmodk is not, however, fault-resilient.

%Research on routing for HPC interconnection neworks
%has looked into fault-resiliency%
%\cite{flich2012survey}\cite{bogdanski2012discovery}\cite{domke2014fail}
%especially when considering large scale clusters
%in which occasional equipment failure becomes unavoidable.
Section~\ref{sec:background} goes into detail
about existing works for fault-resilient fat-tree routing
and their shortcomings.
The proposed fault-resilient routing algorithm
is presented in Section~\ref{sec:algo}.
%alongside some information about our implementation.
Section~\ref{sec:results} then presents results
by comparing our implementation
against routing engines of OpenSM~\cite{opensm}
(those deemed appropriate for routing of degraded fat-trees),
first for routing runtime
and then for congestion risk under degradation.
%Section~\ref{sec:gdmodc} will shortly present a node-type specific variant
Section~\ref{sec:conclusion} summarizes pros and cons
of the proposed technique.

\section{Background}%
\label{sec:background}

Some of the research regarding oblivious fault-resilient routing
focuses on techniques that apply to any connected network%
~\cite{flich2012survey}\cite{domke2014fail};
these topology-agnostic techniques require
full re-routing on topology change.
Some other research explicitly targets degradations
to regular fat-trees~\cite{zahavi2014quasi}\cite{quintin2016transitively};
there are several re-routing strategies for these techniques.
OpenSM's UPDN~\cite{mlx2007updndoc} %\cite{mlx2007updnonline}
and Ftree~\cite{zahavi2010optimized} routing engines
can also be applied from scratch to a degraded fat-tree.
%(provided the list of top switches is given by the user).
PQFT~\cite{zahavi2014quasi} is similar,
though it requires a complete list of faults.
The combination of Dmodk + Ftrnd\_diff~\cite{vigneras2015fault}
available in BXI~FM~\cite{bxi}
is applied in an offline/online manner
(with an iterative list of topology changes
and an up-to-date view of the topology),
the goal being fast reaction to faults
with minimal routing changes.
Fabriscale~\cite{villanueva2015routing} also provides
fast centralized re-routing of fat-trees,
by precomputing alternative routes.

The random operation chosen in Ftrnd\_diff
results in progressive degradation of load balance
and incapacity to return to the original routing
in case of fault recovery.
Ftrnd\_diff does manage to recover rapidly from minor failures;
however large numbers of simultaneous changes
(which happen for example when entire islets are rebooted)
cause computation times to skyrocket in current implementations.
%PQFT and Fabriscale have not been studied here
%because PQFT's implementation is not available yet
%and Fabriscale's algorithm is not presented;
%however the strategies described of moving only invalidated routes
The strategies of PQFT and Fabriscale
which consist in moving only invalidated routes
let one expect somewhat similar
load-balancing issues as with Ftrnd\_diff.
Studies show topology-agnostic routing
outperforms fat-tree-specific routing under sufficient
topology degradations~\cite{bogdanski2012discovery}\cite{domke2014fail}.

\section{Dmodc Description}\label{sec:algo}

The idea behind the fault-resilient algorithm
that we propose in this paper
is to rely on local information like Ftree
(and no topological address),
while %improving load balance and expressed parallelism
using the same closed-form arithmetic operation as Dmodk.
%Some judicious separation of preprocessing and computation
%is required to allow reaching fast computation time.
The aim is fast centralized computation
of routing tables for degraded PGFTs,
providing optimal or well-balanced deterministic routes
even under heavy fabric degradation.
The algorithm begins with a partly sequential preprocessing phase
%described in~\ref{sec:preprocess}
followed by a parallel computation phase.
%described in~\ref{sec:computation}.
%Subsection~\ref{sec:preprocess} describes
%the largely sequential preprocessing phase,
%and Subsection~\ref{sec:computation} describes the following
%embarassingly parallel computation phase.
Links are bi-directional;
notations used in the expressions below
are defined in Table~\ref{tab:notation}.

\subsection{Preprocessing}\label{sec:preprocess}

%Preprocessing generates five results described here.

\subsubsection*{Rank}
Levels and link directions are determined
based on leaf switches being equivalent to the lowest level.
%according to one of multiple corresponding methods
%depending on:
%\begin{itemize}
%	\item whether leaf switches are equivalent to the lowest level,
%		%(in which case ranking can be naturally integrated
%		%in the upward phase of Subsection~\ref{sec:cost}),
%		%algorithm~\ref{algo:costdiv}),
%	\item whether the user provides the list of top-level switches,
%		%(in which case ranking can be integrated
%		%in~\ref{sec:cost}
%		%in algorithm~\ref{algo:costdiv}
%		%by flipping propagation directions),
%	\item otherwise top-level switches are automatically determined
%		based on a statistical rule
%		(e.g.\ UPDN's rule\cite{mlx2007updndoc}).
%\end{itemize}

\subsubsection*{Port Groups}
Groups of ports linked to the same switch
are prepared and sorted by universally unique identifier
(UUID, defined at hardware fabrication)
to help with same-destination route coalescing.

%\begin{algorithm}
%\begin{lstlisting}[]
%for switch in switches:
%	switch.portGroups = list()
%	for other in switch.otherSwitches:
%		pg = PortGroup(
%			remote = other,
%			ports = list())
%		for port in switch.ports:
%			if port.remote = other:
%				pg.ports.append(port)
%		pg.ports.sortByPortRank()
%		switch.portGroups.append(pg)
%	switch.portGroups.sortByUUID()
%\end{lstlisting}
%\caption{Prepare and sort port groups.}\label{algo:pg}
%\end{algorithm}

\subsubsection*{Cost}\label{sec:cost}
We define the cost $c_{s,l}$
of a switch $s$ to a leaf switch $l$
to be the minimum number of hops between one another
under up\---down restrictions according to rank,
as defined in Algorithm~\ref{algo:costdiv}.
%Cost is formally defined
%in the following sequence of assignments:

%\begin{align}
%	\forall\ s \in S,\quad \forall\ l\in L,\quad&c_{s,l} \assign \infty \\
%	\forall\ l\in L,\quad&c_{l,l} \assign 0 \\
%	\label{eqn:cost_asc}
%	\forall\ s' \connabove s \in S,\quad \forall\ l \in L,\quad&
%		c_{s',l} \assign \min(c_{s',l},\ c_{s,l}+1) \\
%	\label{eqn:cost_desc}
%	\forall\ s' \connbelow s \in S,\quad \forall\ l \in L,\quad&
%		c_{s',l}  \assign \min(c_{s',l},\ c_{s,l}+1)
%\end{align}
%
%Note that (\ref{eqn:cost_asc}) can be applied once per switch
%by selecting them in ascending order
%(same for (\ref{eqn:cost_desc}) in descending order).

%One efficient method to compute
%all switches' costs to all leaves in a degraded PGFT
%is through an upward propagation phase (level by level)
%followed by a downward propagation phase.
%At each stage, the finite costs of each switch of the considered level
%are incremented and copied over to linked switches in the next level,
%every time the new value is lower than the previous one.
%The complete process using such a method
%has a complexity $\mathcal{O}(n k^{2n-2})$ in write operations
%and $\mathcal{O}(2n k^{2n-1})$ in comparisons
%%and $\mathcal{O}((2n-2)k^{2n-1})$ in comparisons
%for a $k$-ary $n$-tree.

\subsubsection*{Divider}
Dmodc is based on the same arithmetic operation as Dmodk:
it begins with an integer division by the product
of upward arities ($= \#\{s'\ \connabove\ s\}$) of lower levels.
% ($\prod_{k=1}^l w_k$).
To reflect the actual state of the topology,
only local information must be considered;
in turn this operation is based
on the products of up-to-date counts of upswitches,
as defined in Algorithm~\ref{algo:costdiv}.
%From every switch $s$, every down path corresponds to a potential value;
%we combine them into a single value $\Pi_s$
%(the divider of $s$):
%
%\begin{align}
%	\forall\ l \in L,\quad&\Pi_l \assign 1 \\
%	\forall\ s \in S \setminus L,\quad&
%		\Pi_s \assign \max_{s'\connbelow s}
%		(\Pi_{s'}\times\#\{s''\connabove s'\})
%\end{align}

\begin{algorithm}
	\DontPrintSemicolon\SetVlineSkip{2pt}
	\ForEach{$s \in S$}{
		\lForEach{$l \in L$}{
			$c_{s,l} \assign \infty$
		}
		$\Pi_s \assign 1$ \;
	}
	\lForEach{$l \in L$}{
		$c_{l,l} \assign 0$
	}
	\ForEach{$s \in S$ going upwards}{
		$\pi \assign \Pi_s \times
			\#\{s'\ \connabove\ s\}$ \;
		\ForEach{$s'\ \connabove\ s$}{
			\ForEach{$l \in L\ |\ c_{s,l}+1 < c_{s',l}$}{
				$c_{s',l}\assign c_{s,l} + 1$ \;
			}
			\lIf{$\pi > \Pi_{s'}$}{
				$\Pi_{s'} \assign \pi$\hfill($\max$ reduction)
			}
		}
	}
	\ForEach{$s \in S \setminus L$ going downwards}{
		\ForEach{$s'\ \connbelow\ s$}{
			\ForEach{$l \in L\ |\ c_{s,l} + 1 < c_{s',l}$}{
				$c_{s',l}\assign c_{s,l} + 1$ \;
			}
		}
	}
\caption{Compute costs and dividers.}\label{algo:costdiv}
\end{algorithm}

In a full PGFT, $\#\{s'\ \connabove\ s\}$ values
are constant throughout each level,
and individual values are decreased following faults;
the $\max$ reduction accordingly helps
keep load repartition stable under few faults.
This choice of reduction was only compared with one
using the first downward path and showed little to no change
in route quality under random degradation.

%This step can be implemented alongside cost computation
%by multiplicative propagation of the number of uplinks upwards,
%in which all dividers are initialized to 1
%and values are written only if they are strictly bigger
%than the existing one.
%The corresponding complexity
%is $\mathcal{O}(2k^{n-1})$ in write operations
%and $\mathcal{O}(n k^n)$ in comparisons
%%and $\mathcal{O}((n-1)k^n)$ in comparisons
%for a $k$-ary $n$-tree.
%

%\begin{algorithm}
%\begin{lstlisting}[morekeywords={propagate}]
%# Initialize costs and dividers
%for switch in switches:
%	for leaf in leaves:
%		switch.cost[leaf] = MAX
%	switch.divider = 1
%for leaf in leaves:
%	leaf.cost[leaf] = 0
%# Propagate cost if it is better
%def propagate(switch, other):
%	for leaf in leaves:
%		if other.cost[leaf] - 1 > switch.cost[leaf]:
%			other.cost[leaf] = switch.cost[leaf] + 1
%# Upward propagation
%for level in levels:
%	for switch in level:
%		upDiv = switch.divider * len(switch.upSwitches)
%		for up in switch.upSwitches:
%			propagate(switch, up)
%			if up.divider < upDiv:
%				up.divider = upDiv
%# Downward propagation
%for level in reversed(levels):
%	for switch in level:
%		for down in switch.downSwitches:
%			propagate(switch, down)
%\end{lstlisting}
%\caption{Compute all costs and dividers.}\label{algo:costdiv}
%\end{algorithm}

\begin{table}
\centering
\caption{Notations used in expressions.}\label{tab:notation}
	\begin{tabular}{r@{\hskip 6pt}l}
	$S$         & is the set of switches, \\
	$L$         & is the set of leaf switches ($L\subset S$), \\
	$U(s)$      & is the UUID of switch $s$, \\
	$N$         & is the set of nodes, \\
	$\lambda_n$ & is the (only) leaf switch connected to node $n$ ($\lambda_n\in L$) \\
	%$\conn$     & denotes the existence of a link between its operands, \\
	$\connabove,\connbelow$ & respectively denote downlinks and uplinks
		(according to rank), \\
	$G_s$       & is the ordered list of port groups of switch $s$, \\
	$\Omega_g$  & is the switch connected to port group $g$, \\
	$\#$        & denotes cardinality, in number of port groups or of ports. \\
	$c_{s,l}$   & is the cost of switch $s$ to leaf switch $l$, \\
	$\Pi_s$     & is the divider of switch $s$, \\
	$t_n$       & is the topological node identifier (NID) of node $n$, \\
	\end{tabular}
\end{table}

\subsubsection*{Topological NID}\label{sec:toponid}

The arithmetic nature of Dmodc guarantees load-balancing
only if NIDs (on which the modulo operation is applied)
are topologically contiguous.
We explicitly determine each node's topological NID
using previously computed costs
in Algorithm~\ref{algo:toponid}.

%and a UUID-based global ordering:
%
%\begin{align}
%	\label{eqn:toponid_Gamma}
%	\forall\ l,l'\in L,\quad
%		&\Gamma_{l,l'}
%		=\{l''\in L\ |\ c_{l,l''}<c_{l,l'}\} \\
%	\label{eqn:toponid_gamma}
%	\forall\ l,l'\in L,\quad
%		&\gamma_{l,l'}
%		=\min_{l''\in\Gamma_{l,l'}}(U(l'')) \\
%	\label{eqn:toponid_range}
%	\forall\ n\in N,\quad &t_n\in\interval[open right]{0}{\#N} \\
%	\label{eqn:toponid_order}
%	\forall\ n,n'\in N,\quad
%		&\gamma_{\lambda_n,\lambda_{n'}} < \gamma_{\lambda_{n'},\lambda_n}
%		\Rightarrow t_n<t_{n'}
%\end{align}
%
%Note that (\ref{eqn:toponid_Gamma}) provides
%a cost-based definition of relative groups,
%from which (\ref{eqn:toponid_gamma})
%defines a group-wise UUID\@.
%Using this object, (\ref{eqn:toponid_range}) and (\ref{eqn:toponid_order})
%define a total ordering of topological NIDs
%between nodes of different leaves.
%Nodes linked to the same leaf switch are ordered according
%to their port rank.

\begin{algorithm}
	\DontPrintSemicolon\SetVlineSkip{2pt}
	$t \assign 0$ \;
	$X \assign L$ sorted by UUIDs \;
	\While{$X \ne \text{\O}$}{
		$l \assign X_0$ \;
		$\mu \assign \min_{l'\in X\setminus\{l\}}(c_{l,l'})$ \;
		\ForEach{$l' \in X\ |\ c_{l,l'} \le \mu$}{
			\ForEach{$n\ \connbelow\ l'$ in port rank order}{
				$t_n \assign t$ \;
				$t \assign t + 1$ \;
				$X \assign X \setminus \{l'\} $ \;
			}
		}
	}
\caption{Compute topological NIDs.}\label{algo:toponid}
\end{algorithm}

Dmodc can provide optimal results
for shift permutation communication patterns
which respect such an ordering,
otherwise one could expect results similar
to those of random permutation communication patterns.

%\begin{algorithm}
%\begin{lstlisting}[morekeywords={addTopoNIDs}]
%topoNID = 0
%def addTopoNIDs(leaf):
%	nodes = leaf.nodes
%	nodes.sortByPortRank()
%	for node in nodes:
%		port.remote.topoNID = topoNID
%		topoNID = topoNID + 1
%todo = sortedByUUID(leaves)
%curr = todo[0]
%while todo n= $\text{\O}$:
%	addTopoNIDs(curr)
%	todo.remove(curr)
%	mincost = min((curr.cost[leaf] for leaf in todo))
%	for leaf in todo:
%		if curr.cost[leaf] == mincost:
%			curr = leaf
%			break
%\end{lstlisting}
%\caption{Compute topological NIDs.}\label{algo:toponid}
%\end{algorithm}

\begin{figure*}
	\includegraphics[width=\textwidth]{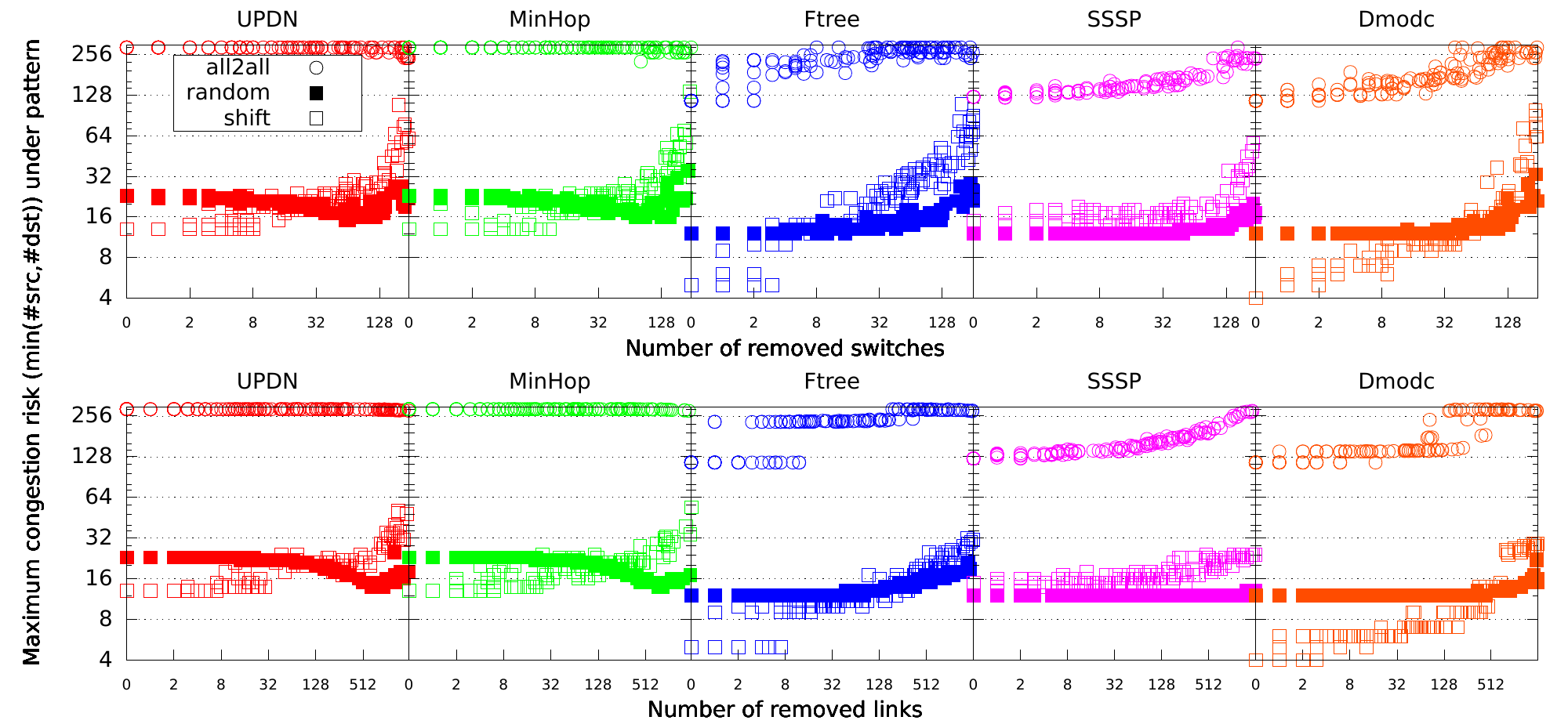}
	\caption{Maximum congestion risk
		in a 8640 node PGFT (with blocking factor of 4)
		under random topology degradation
		(in log\---log scale; lower is better).}\label{fig:perf}
\end{figure*}

\subsection{Routes Computation}\label{sec:computation}

The deterministic output port $p_{s,d}$
and alternative output ports $P_{s,d}$ of every switch $s$
for every destination $d\in N$
(not directly linked to $s$)
are selected with a closed-form operation
based on the results previously determined.
First, port groups leading \textit{closer} to $\lambda_d$
are selected in (\ref{eqn:closer}),
setting corresponding alternative output ports in (\ref{eqn:dmodc_P}):

\begin{align}
	C_{s,\lambda_d} \assign&\left\{ g \in G_s
	\ |\ c_{\Omega_g,\lambda_d} < c_{s,\lambda_d} \right\}\label{eqn:closer}\\
	P_{s,d} \assign\ &\{p \in g\ |\ g \in C_{s,\lambda_d}\}\label{eqn:dmodc_P}
\end{align}

Selected port groups $C_{s,\lambda_d}[i]\ \forall\ i\in \#C_{s,\lambda_d}$
are ordered by UUID of their remote switch.
From this, the output port group is chosen in (\ref{eqn:dmodc_pg})
and the port within that group in (\ref{eqn:dmodc_p}):

\begin{align}
	g_{s,d} \assign\ &C_{s,\lambda_d}[\left\lfloor \frac{t_d}{\Pi_s} \right\rfloor
	\bmod \#C_{s,\lambda_d}]\label{eqn:dmodc_pg} \\
	p_{s,d} \assign\ &
	g_{s,d}[\left\lfloor \frac{t_d}{\Pi_s\times \#C_{s,\lambda_d}} \right\rfloor
	\bmod \#g_{s,d}]\label{eqn:dmodc_p}
\end{align}

%Deterministic routes are computed with algorithm~\ref{algo:dmodc}.

%Alternative tables are determined in (\ref{eqn:dmodc_P}).

%If alternative tables are required for adaptive routing,
%each one can be simply computed as the set of ports in $C_{s,\lambda_d}$
%for its destination $d$.

%(\texttt{c} is the list of switches \textbf{closer}
%to \texttt{l} than \texttt{sw}, ordered by UUID)
%\begin{algorithm}
%\begin{lstlisting}
%for switch in switches:
%	for dest in nodes:
%		switch.ART[dest.NID] = $\text{\O}$
%		leaf = dest.remote
%		# Build list of port groups closer to dest
%		closer = list()
%		for pg in switch.portGroups:
%			if pg.remote.cost[leaf] < switch.cost[leaf]:
%				closer.append(pg)
%				# Build alternative routing table
%				switch.ART[dest.NID].extend(pg.ports)
%		# Choose port group using dmodc formula
%		c = len(closer)
%		pg = closer[$\left\lfloor\frac{\text{dest.topoNID}}{\text{switch.divider}}\right\rfloor$ mod c]
%		# Choose port within port group
%		p = pg.ports[$\left\lfloor\frac{\text{dest.topoNID}}{\text{switch.divider}\times\text{c}}\right\rfloor$ mod len(pg.ports)]
%		# Update linear forwarding table (LFT)
%		switch.LFT[dest.NID] = p
%\end{lstlisting}
%\caption{Compute all routing tables}\label{algo:dmodc}
%\end{algorithm}

\section{Results}\label{sec:results}

\subsection*{Validity}

Routing is valid for degraded PGFTs
if and only if the cost of every leaf switch
to every other leaf switch is finite:
this reflects every node pair having an up\---down path.
Our implementation includes a pass through all leaf switch pairs
to verify this condition.
The up\---down path restriction is sufficient
to guarantee deadlock-freedom within degraded PGFTs~%
\cite{quintin2016transitively}.

%Experimentation results are given in table~\ref{tab:times}.
%
%\begin{table}
%	\centering
%	\caption{Average execution times on a 2.5GHz 48-core Xeon CPU.}\label{tab:times}
%	\begin{tabular}{|r|r|r|r|}
%		\hline
%		Topology & Preprocess & Deterministic & Alternative \\
%		\hline
%		\hline
%		8640 node PGFT & 14ms & 10ms & 7ms \\
%		65535 node RLFT & 137ms & 710ms & \\
%		\hline
%	\end{tabular}
%\end{table}

\subsection*{Congestion Risk}

Random degradation is simulated using hundreds of throws
for each considered routing algorithm
and type of equipment to degrade (switches or links).
The integer amount of equipment
$a\in\interval[open right]{0}{2^m}$
to remove at each throw is chosen
using a shifted log-uniform distribution.
This distribution is chosen to test degradation
uniformly across multiple scales
and include non-degraded tests;
it is defined in the following formula
using uniform number generator $u() \in \interval{0}{1}$:

\begin{displaymath}
	a \assign \left\lfloor 2^{m\times u()} - 1 \right\rfloor
\end{displaymath}

The chosen amount of equipment is then randomly removed
from the complete topology.
The resulting degraded (or complete) topology is routed at this point,
and linear forwarding tables are dumped for analysis.

Evaluation of these tables is performed
using static analysis of metrics representing maximum congestion risk
for three communication patterns:
all-to-all (A2A), random permutation (RP), shift permutation (SP).
The congestion risk metric consists of counting $\min(\#srcs,\#dsts)$
for all routes of the corresponding pattern;
this approximates network-caused congestion risk~\cite{rodriguez2009exploring}.
For A2A, the maximum congestion risk (throughout all ports)
is the only value kept.
RP consists of computing the maximum congestion risk
for 1000 random permutations
and keeping the median value.
%(The number of permutations was determined experimentally
%to result in stable maximum congestion risks:
($\sigma=0.96$ for 100 RP samples
in the case of Figure~\ref{fig:perf}
with 256 randomly removed switches routed with Ftree.)
SP consists of computing the maximum congestion risk
for all ($\#N - 1$) shift permutations
and keeping the maximum value.
(Shifts are based on the same node ordering
which OpenSM's Ftree follows internally
in order for quality comparison to be fair.)
Such simplified performance models
faithfully reflect comparative behaviour~\cite{faizian2017throughput},
though the absolute values measured
are not good estimators of real throughput.
Note that virtual channels potentially required by other algorithms
are not taken into account in this analysis.

Congestion risk results are shown in Figure~\ref{fig:perf}.
When considering existing routing algorithms,
Ftree provides the best performance for complete PGFTs
(especially regarding SP for which
the maximum congestion risk approaches theoretical optimal),
but SSSP provides better stability under massive degradation,
confirming results of the studies mentioned in Section~\ref{sec:background}.
UPDN and MinHop provide visually identical results in this analysis:
in fact in a full PGFT they are equivalent
and vary only slightly under degradation.
They both provide comparatively poor results for SP and A2A
throughout the observed scale,
however for RP they surprisingly improve significantly
under massive degradation.

Dmodc provides minimal congestion risk
throughout the considered range of degradations
when compared with existing oblivious algorithms.
In particular, it is even more stable than Ftree for SP
under minimal degradation
and nearly as stable as SSSP for A2A and RP
under massive degradation.

%If one wants to minimize size of updates
%without interstate knowledge,
%one strategy is to pad NIDs.
%This has not been studied in practice yet.

%\section{gDmodc}
%
%Reindexing à la gDmodk is compatible with Dmodc
%and should yield similar results in a more resilient way.
%This has not been studied in practice yet.

\begin{figure}
	\includegraphics[width=\linewidth]{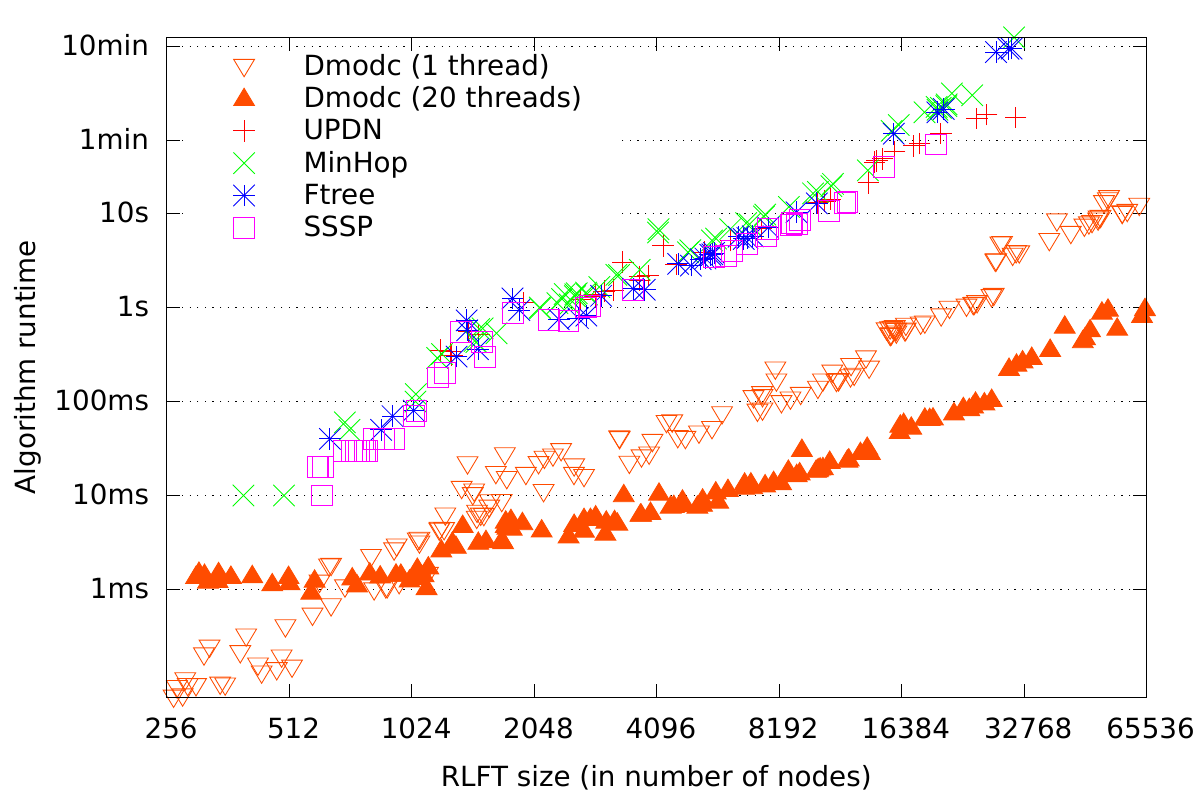}
	\caption{Algorithm runtime %with 1 and 20 threads
		on an Intel Xeon E5{-}2680 v3 @~2.50GHz
		(in log\---log scale; lower is better).}%
	\label{fig:times}
\end{figure}

\subsection*{Runtime}

On our production implementation coded in C99,
computation of cost, divider, topological NIDs, and routes
%preprocessing steps~\ref{sec:cost} to~\ref{sec:toponid} and route computation
are spread over POSIX threads fetching work with a switch-level granularity.
Figure~\ref{fig:times} reports complete algorithm execution time
%(including alternative table computation),
alongside OpenSM (version 3.3.21) routing times
(measured by adding timers in the source code)
running on the same machine.
Note that local erraticness is partly due
to our RLFT construction process
for which the number of resulting switches
is not monotonic with the number of requested nodes.

For clusters ranging up to many tens of thousands of nodes,
Dmodc provides fast enough re-routing
for a centralized fabric manager to react to faults
before applications are interrupted.

\section{Conclusion}\label{sec:conclusion}

The simulation results in Section~\ref{sec:results}
show that Dmodc provides high-quality
centralized fault-resilient routing for PGFTs
at a fraction of the runtime of existing algorithms,
without relying on partial re-routing.
Dmodc is also applicable to non-PGFT fat-tree-like topologies
but with lower quality load balancing.
As defined here, no effort has been made to minimize
size of updates to be uploaded to switches throughout the fabric.

This algorithm is implemented inside BXI~FM
and has been succesfully deployed
to an 8490 node PGFT production topology
in which it helps provide fault-resiliency
even when faced with thousands of simultaneous changes.

\section*{Acknowledgements}
This research has been undertaken
under a cooperation between CEA and Atos.
with the goal of co-designing
extreme computing solutions.
Atos thanks CEA for all their inputs
that were very valuable for this research.
This research was partly funded by a grant
of Programme des Investissements d'Avenir.
This work has been jointly supported
by the Spanish Ministry of Science, Innovation and Universities
under the project RTI2018{-}098156{-}B{-}C52
and by JCCM under project SBPLY/17/180501/000498.
BXI development was also part of ELCI,
the French FSN (Fond pour la Société Numérique)
cooperative project that associates
academic and industrial partners
to design and provide software components
for new generations of HPC datacenters.

%\IEEEtriggeratref{9}

\bibliographystyle{IEEEtran}
\bibliography{dmodc}

\end{document}